\documentclass[aip, rsi, amsmath, amssymb, reprint]{revtex4-1}

\usepackage{graphicx}
\usepackage{dcolumn}
\usepackage{bm}
\usepackage[utf8]{inputenc}
\usepackage[T1]{fontenc}
\usepackage{mathptmx}
\usepackage{gensymb}
\usepackage{upgreek}
\usepackage{xcolor}

\newcommand{\comment}[1]{}
\usepackage{titlesec}
\titleformat{\section}[hang]{\small\bfseries\sffamily}{\thesection.}{0.5em}{\MakeUppercase}
\titlespacing{\section}{0pc}{1pc}{0.2pc}

\begin{document}
\preprint{APL}
\title{Superconducting nanowire single-photon detectors \\ fabricated  from atomic-layer-deposited NbN}

\author{Risheng Cheng}
\author{Sihao Wang}
\author{Hong X. Tang}
\email{hong.tang@yale.edu}
\affiliation{Department of Electrical Engineering, Yale University, New Haven, CT 06511, USA}
\date{\today}

\begin{abstract}
High-quality ultra-thin films of niobium nitride (NbN) are developed by plasma-enhanced atomic layer deposition (PEALD) technique. Superconducting nanowire single-photon detectors (SNSPDs) patterned from this material exhibit high switching currents and saturated internal efficiencies over a broad bias range at 1550\,nm telecommunication wavelength. Statistical analyses on hundreds of fabricated devices show near-unity throughput yield due to exceptional homogeneity of the films. The ALD-NbN material represents an ideal superconducting material for fabricating large single-photon detector arrays combining high efficiency, low jitter, low dark counts.

\end{abstract}
\maketitle


Superconducting nanowire single-photon detectors (SNSPDs)\cite{natarajan_2012_SNSPD_review, goltsman_2001_first_SNSPD} have gained widespread attention due to their excellent performances, including high efficiency\cite{marsili_2013_93p_efficiency,simit_2017_92p_nbn_detector,delft_2017_92p_nbn_detector_Delft,nist_2019_95p_mosi_detector}, fast speed\cite{simit_2019_16_pixel_detector,pernice_2016_1D_PhC_detector}, exceptional timing jitter\cite{jpl_2018_low_jitter,Delft_2018_10ps_jitter_detector,JPL_2019_MoSi_detector_jitter} and ultra-low dark count rates\cite{schuck_2013_mHz_dark_count,NTT_2015_ultimate_darkcounts}. In addition, their suitability for on-chip integration with various nanophotonics circuits \cite{Fiore_2011_waveguide_snspd,pernice_2012_waveguide_SNSPD,schuck_2013_nbtin_detector_sin_waveguide,Fiore_2015_GaAs_SNSPD,pernice_2015_waveguide_snspd,pernice_2015_snspd_diamond,berggren_2015_on_chip_detector,hadfield_2016_MoSi_waveguide_detector,pernice_2018_waveguide_snspd_review,italy_2019_amplitude_multiplex} as well as their photon number resolving  \cite{fiore_2008_pnd,cheng_2013_SND,berggren_2018_scalable_detector} and spectral \cite{pernice_2017_spectrally_multiplexed_snspd,cheng_2019_broadband_spectrometer,kit_2019_phc_spectrometer} resolving capability render them an ideal choice for applications in quantum optics, quantum communications and quantum information processing\cite{yamamoto_2007_quantum_key_snspd,liao_2017_satellite,wang_2018_multidimensional_quantum}.      

\begin{figure*}[!ht]
\centering\includegraphics[width=17cm]{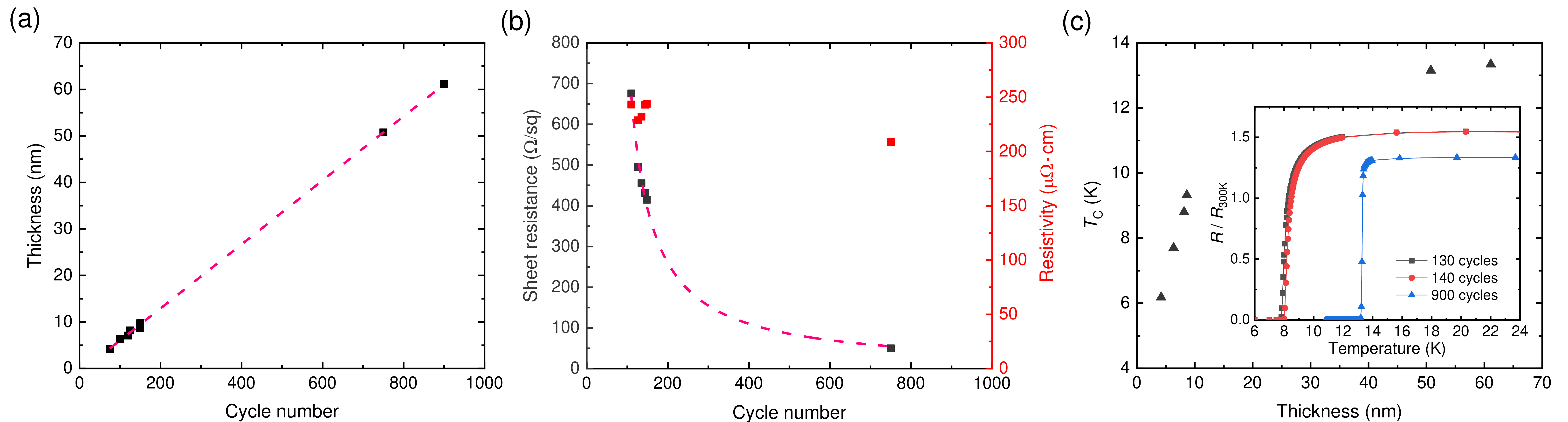}
\caption{
(a) Measured NbN film thickness versus the ALD cycle number. The dashed line is a linear fit, whose slope indicates the deposition rate of $0.68$\,\AA/cycle. 
(b) Measured sheet resistance ($R_\mathrm{sheet}$) and calculated resistivity of the NbN films versus the ALD cycle number ($x$). The dashed line represents a fit by the equation $R_\mathrm{sheet} = B/(x-a)$ with $B$ and $a$ as free fitting parameters.
(c) $T_\mathrm{c}$ of the NbN films versus the thickness.
Inset: resistance of three NbN films versus temperature showing the superconducting transition. The resistance is normalized by the room temperature value $R_\mathrm{300K}$. }
\label{Tc_resistance}
\end{figure*}

The SNSPD is typically a narrow nanowire (20\,nm\,-\,150\,nm width) patterned from an ultra-thin superconducting film (3-10\,nm thickness) for absorbing incident photons. So far, two main classes of superconducting materials have been utilized to fabricate high-efficiency SNSPDs: (1) poly-crystalline nitride superconductors such as NbN\cite{simit_2017_92p_nbn_detector,pernice_2012_waveguide_SNSPD,dauler_2013_nbn_detector_array} and NbTiN\cite{schuck_2013_nbtin_detector_sin_waveguide,cheng_2016_self_aligned_detector,cheng_2017_multiple_SNAP,delft_2017_92p_nbn_detector_Delft,nict_2017_nbtin_snap}; (2) amorphous alloy superconductors, such as WSi\cite{nist_2011_first_wsi_detector,marsili_2013_93p_efficiency},  MoSi\cite{goltsman_2014_mosi_detector,hadfield_2016_MoSi_waveguide_detector,switzerland_2018_mosi_detector,JPL_2019_MoSi_detector_jitter,nist_2019_95p_mosi_detector} and MoGe\cite{nist_2014_moge_detector}. Each of these two classes of materials has their advantages and drawbacks in photon detection. The amorphous films have smaller superconducting energy gap and lower electron density, which tends to create larger photon-excited hot spots in the nanowires and thus results in better saturated internal efficiency. The structural homogeneity due to the absence of grain boundaries in these films further enables the fabrication of large-area detector arrays without suffering from serious constrictions\cite{jpl_2017_large_area_wsi_detector,jpl_019_kilopixel_snspd}. The drawback is their relatively lower superconducting transition temperature $T_\mathrm{c}$ (<\,5\,K for thin films) which sets their operation temperature below 2.5\,K in order to achieve saturated efficiency at 1550\,nm wavelength\cite{nist_2014_2.5k_wsi_detector}. Poly-crystalline Nb(Ti)N-based detectors, on the other hand, have higher $T_\mathrm{c}$, higher critical current,  relatively improved jitter performance and more immunity to latching at high operation speed due to their shorter hot spot relaxation time\cite{nist_2016_hotspot_detector,pernice_2017_hotspot_snspd,simit_2018_hotspot_snspd}. Consequently, Nb(Ti)N was exploited in the recent demonstration of GHz-counting-rate detectors\cite{pernice_2016_1D_PhC_detector,simit_2019_16_pixel_detector} and sub-3\,ps timing jitter\cite{jpl_2018_low_jitter}.    

In this Letter, we show our development of high-quality NbN thin films by plasma-enhanced atomic layer deposition (PEALD) and their applications in SNSPDs fabrication. The fabricated detectors demonstrate broad saturated plateaus in the efficiency curves that are comparable with amorphous detectors, while simultaneously maintaining high switching currents. Statistical measurements over hundreds of detectors show a close-to-unity throughput yield of large-area detectors, indicating the high homogeneity of our films. ALD-NbN shown here provides a scalable material platform for realizing large array of single-photon detectors with saturated efficiency and low timing jitter.


The NbN films are prepared by the Ultratech/CNT Fiji ALD system using (tert-butylimido)-tris(diethylamido)-niobium(V) (TBTDEN) as the niobium precursor and mixed N\textsubscript{2}/H\textsubscript{2} as the plasma gas. In each cycle of deposition, the substrate is first exposed to the precursor to form a monolayer of TBTDEN on the surface, which is later reduced to NbN by reacting with the plasma. Ar purging is performed prior to each plasma step to remove residual TBTDEN. By alternating these steps,  NbN film is deposited on the substrate with atomic level thickness control. The best results combining the highest $T_\mathrm{c}$ as well as the lowest resistivity are achieved at the substrate temperature of 300\,\degree C and 300\,W plasma power. More details on the preparation process including other optimized parameters could be found in Ref\cite{sowa_2017_ald_nbn}.  

Figure\,1(a) plots the measured NbN film thickness with respect to the number of deposition cycles. All the NbN films are deposited on thermally oxidized Si chips with 330\,nm-thick LPCVD-grown SiN$_x$ on top for future photonic circuit integration of the SNSPDs\cite{schuck_2013_nbtin_detector_sin_waveguide,cheng_2019_broadband_spectrometer}. The thickness is measured by a spectroscopic ellipsometer (J.A. Woollam M-2000). The slope extracted from the linear fit (dashed line) indicates a deposition rate $DR$ = 0.68\,$\pm$\,0.01\,\AA/cycle. Figure\,1(b) shows the sheet resistance measured by the four probe and van der Pauw method on the unpatterned bare NbN films as a function of the cycle number. The dashed line represents a fit to the experimental data using the equation $R_\mathrm{sheet} = B/(x-a)$, where $R_\mathrm{sheet}$ and $x$ denote the sheet resistance and the cycle number, respectively. The best fitting is done by the set of free parameters $B = 38178\,\upOmega/\mathrm{sq}$ and $a = 52.9$. The non-zero value of $a$ suggests the existence of ``dead cycles'', which does not contribute to the conductance of the film. Based on the obtained $DR$ in Fig.\,1(a), we estimate the corresponding thickness to the ``dead cycles'' to be approximately 3.6\,nm, which is significantly larger than the sputtered NbN films shown in other work\cite{berggren_2019_nbn_thickness}. The underlying mechanism is subject to further investigation. We also show in Fig.1\,(b) the calculated resistivity by multiplying $R_\mathrm{sheet}$ with the effective thickness $(x-a)\times DR$. We observe consistent resistivity of around 240\,$\upmu\upOmega\cdot\mathrm{cm}$ for the thin films (<10\,nm thickness), which is slightly larger than 210\,$\upmu\upOmega\cdot\mathrm{cm}$ obtained from thicker films (>50\,nm thickness).    

In order to investigate the superconducting property of the NbN films, multiple NbN chips are cooled down in a physical property measurement system (Quantum Design PPMS DynaCool) to record the temperature dependence of the resistance. Figure\,2(c) shows the film thickness dependence of $T_\mathrm{c}$, which is defined as the temperature where the resistance of the film is dropped to 50\% of $R_\mathrm{20K}$ the resistance measured at 20\,K. As expected, higher $T_\mathrm{c}$ is observed for thicker film, and it tends to saturate at 13.3\,K with 61\,nm thickness (900 cycles). 
The inset of Fig.\,2(c) shows the zoomed-in view of the resistance versus temperature curves around the superconducting transition region. The very sharp transition in the 900-cycle film with only 0.15\,K transition width (90\% to 10\% of $R_\mathrm{20K}$) suggests high homogeneity and uniformity of the deposited film. The 130-cycle and 140-cycle films that are later fabricated into nanowire detectors exhibit reduced $T_\mathrm{c}$ around 8\,K and broader transition with 1.7\,K width. 

\begin{figure}[!htbp]
\centering\includegraphics[width=8.5cm]{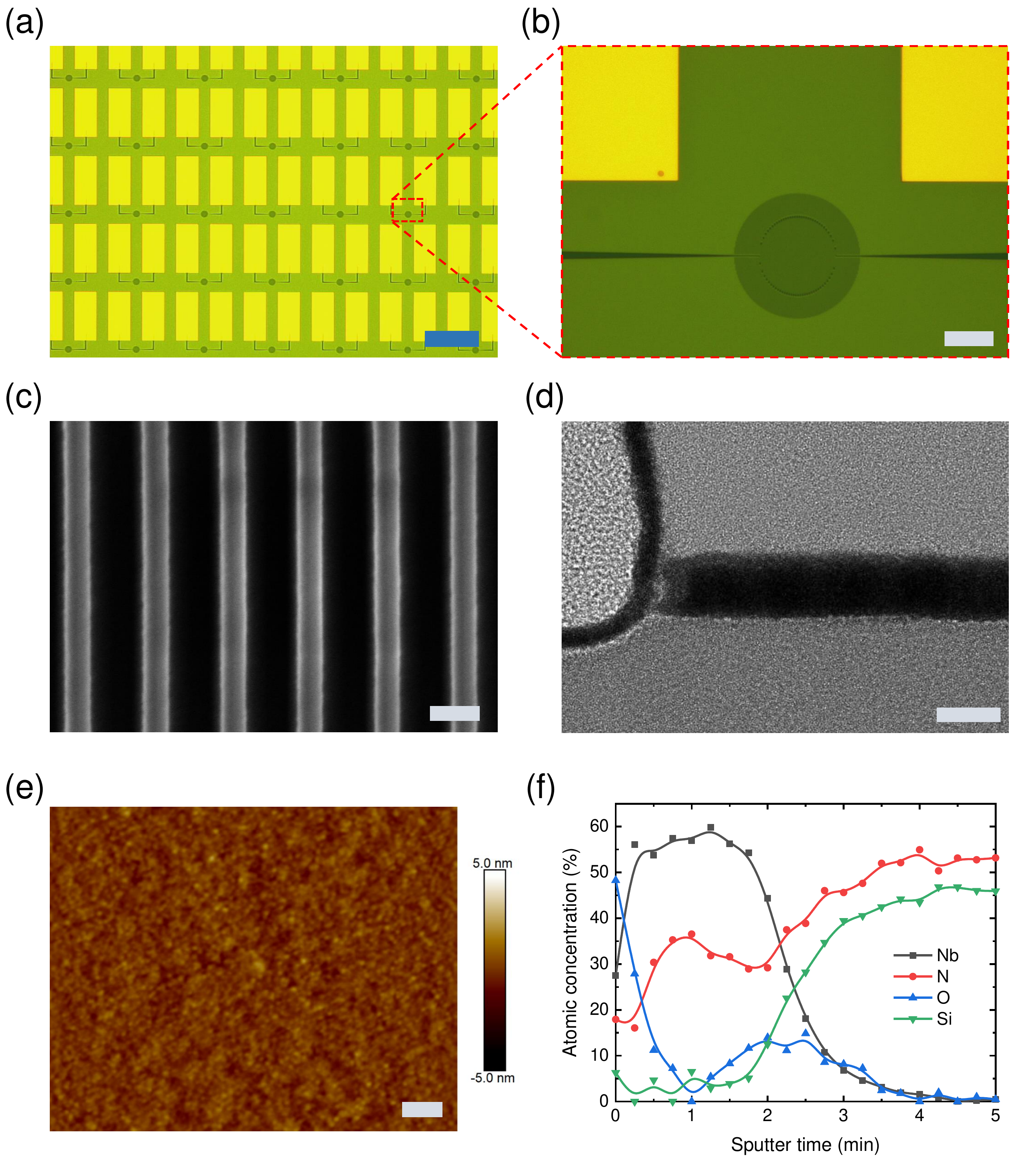}
\caption{
(a) Optical micrograph of fabricated SNSPD array. Scale bar, 200\,$\upmu$m.
(b) Close-up view of the nanowire detection area of an SNSPD. Scale bar, 10\,$\upmu$m.
(c) Close-up SEM image of an SNSPD with 50\,nm-width nanowires. The pitch of the nanowires are kept three times of the width. Scale bar, 100\,nm.
(d) TEM image taken at the edge of the nanowire cross-section patterned from the 140-cycle NbN film. The HSQ mask is left on top after fabrication. Scale bar, 10\,nm.
(e) AFM image taken on the bare 140-cycle NbN film prior to the device patterning. Scale bar, 100\,nm.
(f) XPS measurement results of the 140-cycle NbN film showing the depth profile of the atomic concentration for multiple concerned elements. 
}
\label{device_image}
\end{figure}

\begin{figure}[!htbp]
\centering\includegraphics[width=8cm]{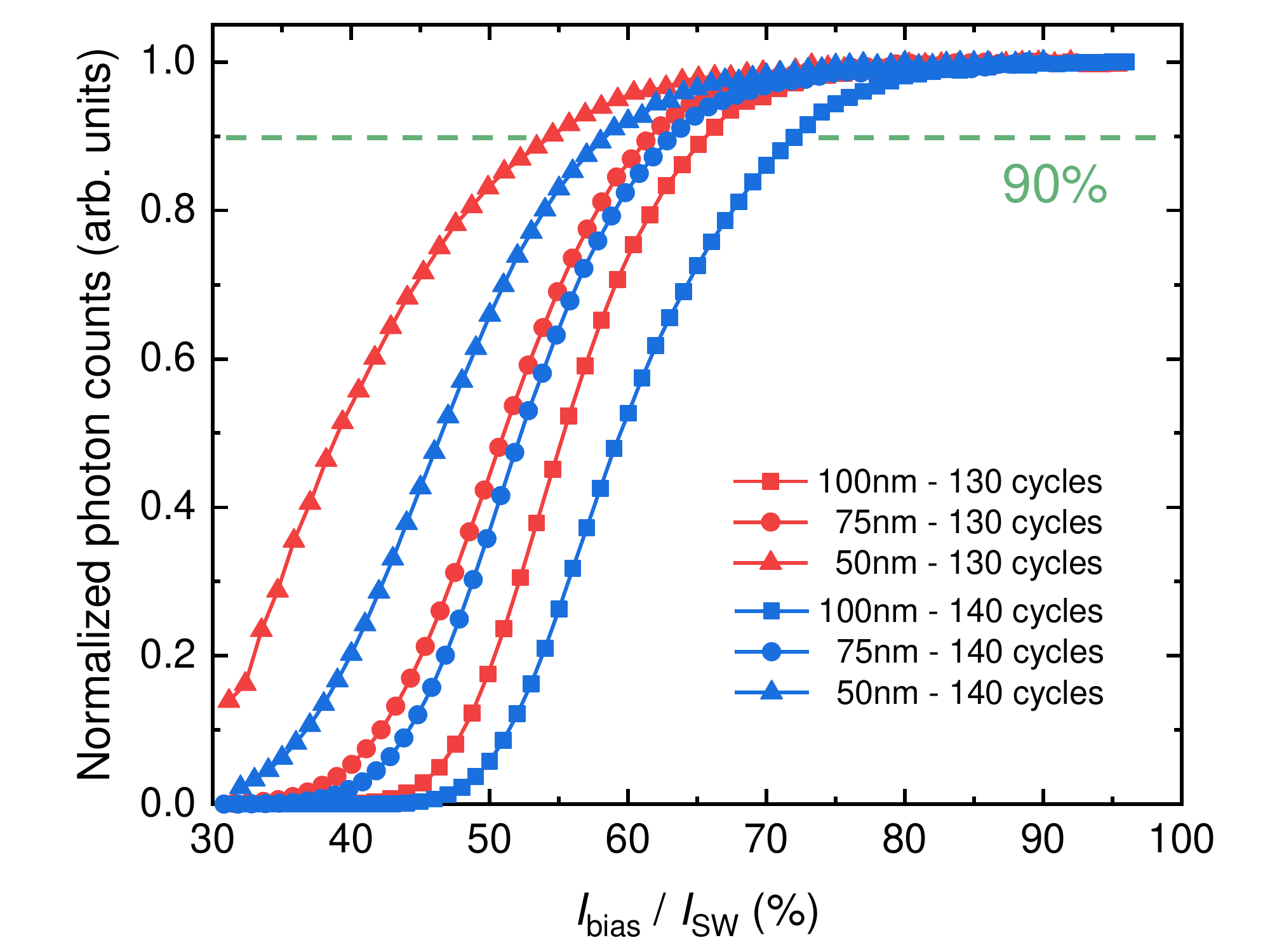}
\caption{
Normalized photon counts versus the relative bias current ($I_\mathrm{bias}/I_\mathrm{SW}$) for SNSPDs of varying width and thickness. The saturation current $I_\mathrm{sat}$ is defined as the current where 90\% of the maximum counting rates is reached.    
}
\label{efficiency}
\end{figure}

\begin{figure*}[!htbp]
\centering\includegraphics[width=13cm]{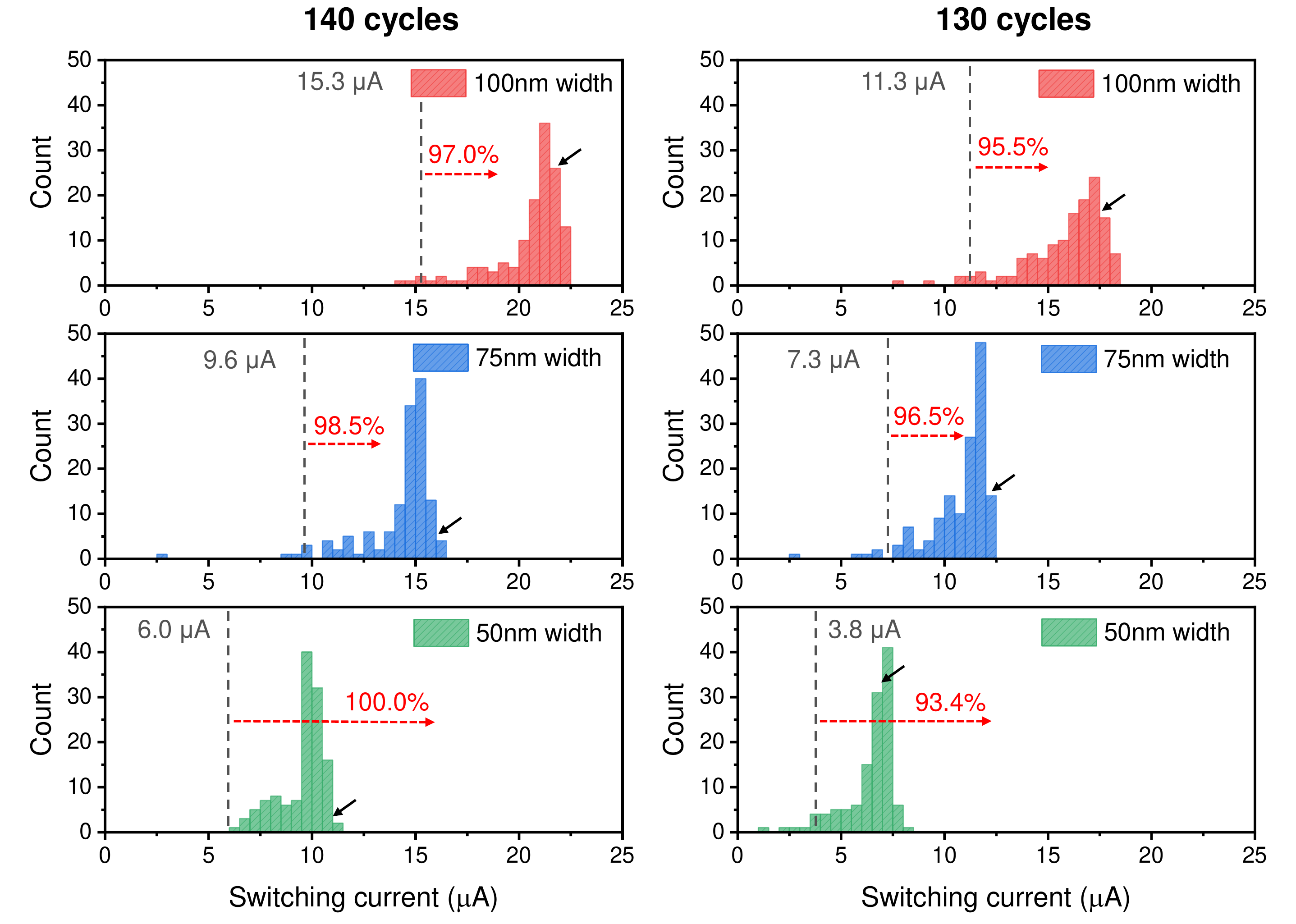}
\caption{
Histogram of $I_\mathrm{SW}$ measured from SNSPDs of varying  width and thickness. The gray dashed lines represent $I_\mathrm{sat}$ for each types of devices, and the calculated throughput yields are shown on the red dashed arrows.  $I_\mathrm{SW}$ of the reference detectors shown in Fig.\,3 are marked using the black solid arrows.    
}
\label{histogram}
\end{figure*}

We fabricate SNSPDs by patterning the 130-cycle and 140-cycle NbN films deposited on SiN$_x$ chips. We define superconducting nanowires by the exposure of negative-tone 6\% hydrogen silsesquioxane (HSQ) resist using high-resolution ($100\,\mathrm{kV}$) electron-beam lithography (Raith EBPG5200) and the subsequent development in tetramethylammonium hydroxide (TMAH)-based developer MF-312. In a second electron-beam lithography step, contact electrodes are defined using double-layer polymethyl methacrylate (PMMA) positive-tone resist. After the development in the mixture of methyl isobutyl ketone (MIBK) and isopropyl alcohol (IPA), we liftoff electron-beam evaporated 10$\,$nm-thick Cr adhesion layer and 100$\,$nm-thick Au in acetone to form the contact pads. Later, the HSQ nanowire pattern is transferred to the NbN layer in a timed reactive-ion etching (RIE) step employing tetrafluoromethane (CF\textsubscript{4}) chemistry.

On each NbN chip, we fabricate a total of 450 SNSPDs with three different designs that respectively employ nanowires of 50\,nm, 75\,nm and 100\,nm nominal width (Fig.\,2(a)). As shown in Fig.\,2(b), the nanowires are meandered into a circular shape of 15\,$\upmu$m diameter to form the active detection area, while the pitch of the nanowires are kept three times of the width. The extra floating nanowires surrounding the active area are for proximity effect correction during the electron-beam exposure, which constitute a larger circle of 25\,$\upmu$m diameter along with the active area. Figure\,2(c) shows the scanning electron micrograph (SEM) image for one of the 50\,nm-width devices. The fabricated nanowires demonstrate high uniformity with less than 5\,nm width variation across the whole active area of the device. In the meantime, the uniformity of the long nanowires are also guaranteed by the ultra-smooth surface of the NbN film. As visualized in Fig.\,2(e), the atomic force micrograph (AFM) image taken on the NbN film prior to the device patterning demonstrate a better than 0.4\,nm root-mean-square surface roughness with negligible difference from the original SiN$_x$ substrate. 

Figure\,2(d) presents the transmission electron micrograph (TEM) image taken at the edge of the nanowire cross-section with the HSQ mask left on top. The total thickness of the nanowire (patterned from 140-cycle NbN film) is approximately 9.5\,nm including the 2\,nm-thick native oxide layer, which could be distinguished from slight color difference. For further investigation of the elemental composition, we perform X-ray photoelectron spectroscopy (XPS) measurement on the unpatterned NbN film. By slowly sputtering the target film using focused Ar ion beam, the depth profile of the atomic concentration for multiple concerned elements could be obtained. As shown in Fig.\,2(f), the existence of the native oxide layer is further confirmed by the high concentration of oxygen observed during the first 0.5 minute sputtering, which decays rapidly as the element analysis is performed deeper into the film. Notably, we observe another oxygen peak at the interface between NbN and SiN$_x$ substrate.  
The existence of this initial layer of oxide leads to change in the property of NbN films and could be the origin of the aforementioned ``dead cycles''. A separate XPS measurement is also carried out on the thicker 900-cycle NbN film after 6 minutes Ar sputtering to completely remove the surface oxide layer and any potential contamination of the film. The element analysis results indicate 60-65\% concentration of Nb and 35-40\% N with negligible presence of other elements.


The detector chips are mounted on a 3-axis stack of Attocube stages inside a closed-cycle refrigerator and cooled down to 1.7 K base temperature. In order to characterize the optical response of the fabricated detectors, 1550 nm continuous wave (CW) laser light is attenuated to the single-photon level and sent to the detector chips via a standard telecommunication fiber (SMF-28) installed in the refrigerator. The detectors are flood-illuminated by fixing the fiber tip far away from the surface of the detector chips, and the diameter of the provided beam spot on the chip is estimated to be around 2 mm. We control the Attocube stages by the LabVIEW program to move the detector chips and make the electrical contact between the RF probes and the gold pads of the detectors. The RF probes are connected to a coaxial cable installed in the refrigerator, and the room-temperature end of the cable is attached to a bias-tee (Mini-Circuits ZFBT-6GW+) to separate the DC bias current and RF output pulses for the detectors. The output pulses are amplified by a low-noise RF amplifier (RF bay LNA-650) and sent to a fast oscilloscope for pulse observation or a pulse counter (PicoQuant PicoHarp 300) for the photon counting measurement.     

All the detectors are screened by an automated measurement program, and several best detectors with the highest switching currents $I_\mathrm{SW}$ are selected for subsequent more detailed optical characterization. Figure\,3 plots the normalized photon counts as a function of the relative bias current ($I_\mathrm{bias}/I_\mathrm{SW}$) for SNSPDs of different width and thickness. The $I_\mathrm{SW}$ of the 100\,nm, 75\,nm and 50\,nm-width nanowires patterned from the 140-cycle (130-cycle) NbN film are 21.9\,$\upmu$A (17.6\,$\upmu$A), 16.3\,$\upmu$A (12.4\,$\upmu$A) and 11.1\,$\upmu$A (6.9\,$\upmu$A), respectively. All the measured detectors demonstrate very broad saturated response plateaus, indicating high internal efficiencies of the detectors. The broad plateaus not only render the detectors less sensitive to the constriction but also enable the operation of the detectors at lower bias region with much lower intrinsic dark counts and without compromising high efficiency. It is worth noting that the broad plateaus shown here are comparable with the amorphous detectors\cite{marsili_2013_93p_efficiency}, while the high $I_\mathrm{SW}$ is not compromised. Our 100\,nm-width detectors made from the 140-cycle NbN film still demonstrate a well-saturated efficiency  with a significantly improved $I_\mathrm{SW}$ larger than 20\,$\upmu$A. This feature eases the fabrication process and also promises faster detectors with low timing jitter.    

The high homogeneity of our NbN films could also be seen in the $I_\mathrm{SW}$ histogram for the detectors of different design visualized in Fig.\,4. The results show highly concentrated distribution of $I_\mathrm{SW}$,  except for just a small number of nanowires showing reduced $I_\mathrm{SW}$ possibly due to the constriction. Since it is extremely time-consuming to perform the optical characterization for all the detectors, we compare the $I_\mathrm{SW}$ with the reference detectors of the same design shown in Fig.\,3 for the estimation of the detector throughput yield. For each type of detectors, we first define a saturation current $I_\mathrm{sat}$ as the current where 90\% of the maximum counting rates is reached (see green dashed lines in Fig.\,3 and gray dashed lines in Fig.\,4). The throughput yield is then calculated as the ratio of the number of the detectors showing $I_\mathrm{SW} > I_\mathrm{sat}$ to the total number. Despite the large areas of the detectors, throughput yields exceeding 93\% are recorded for all types of detectors. In particular, a 100\% yield is demonstrated with 50\,nm-width detectors patterned from the 140-cycle NbN film due to the broader saturation plateau with narrower width and also better uniformity in thickness compared to the slightly thinner 130-cycle film.        

In conclusion, based on PEALD technique, we have prepared high quality and uniform NbN films with $T_\mathrm{c}$ of up to 13.3\,K and atomic level thickness control. The SNSPDs made from this material simultaneously demonstrate broad saturated efficiency plateaus and high $I_\mathrm{SW}$ larger than $20\,\upmu A$. These features not only allow for the operation of the detectors at lower dark counts regime without compromising high efficiencies and low jitters, but also render the detectors less sensitive to the constriction when fabricating large-area detector arrays. In addition, the statistical analysis on the $I_\mathrm{SW}$ histogram measured for hundreds of detectors show a strikingly high device yield approaching unity, which we attribute to the exceptional homogeneity of the film with very few defects. We expect the system efficiencies of our ALD-NbN detectors could be significantly improved by integrating them with optical cavities and careful alignment with optical fibers in the near future\cite{marsili_2013_93p_efficiency,nist_2019_95p_mosi_detector,simit_2017_92p_nbn_detector,delft_2017_92p_nbn_detector_Delft}. We also envision that the ALD-NbN material will play an important role in the fabrication of integrated quantum photonic circuits where on-chip waveguide-integrated SNSPDs are in great demand.  

During the preparation of this manuscript, we became aware of the work by Knehr et al. on SNSPDs made from ALD-NbN \cite{kit_2019_ald_snspd}.

\section*{Acknowledgments}
We acknowledge funding support from DARPA DETECT program through an ARO grant (No: W911NF-16-2-0151), NSF EFRI grant (EFMA-1640959), AFOSR MURI grant (FA95550-15-1-0029), and the Packard Foundation. The authors would like to thank Michael Power, Sean Rinehart, Kelly Woods, Dr. Yong Sun, Dr. Min Li, Dr. Lei Wang and Dr. Michael Rooks for their assistance provided in the film deposition, characterization and device fabrication. The characterization of the NbN films was done at the Yale West Campus Materials Characterization Core (MCC). The fabrication of the devices was done at the Yale School of Engineering \& Applied Science (SEAS) Cleanroom and the Yale Institute for Nanoscience and Quantum Engineering (YINQE).

\def\bibsection{\section*{References}}

%

\end{document}